# Real-Time Pitch/F0 Detection Using Spectrogram Images and Convolutional Neural Networks

Xu Fang Zhao
*Companions and Platforms*
*General Motors Company*
Warren, Michigan, US
Xufang.Zhao@gm.com

Omer Tsimhoni
*Connected Vehicle Experience*
*General Motors Company*
Warren, Michigan, US
omer.tsimhoni@gm.com

*Abstract*—Pitch (also called F0 or fundamental frequency) is a very important voice feature for smart mobility features, such as driver's emotion detection, vehicle personalized profiles, and secured speaker identification. This paper presents a novel approach to detect F0 through Convolutional Neural Networks (CNN) and image processing techniques to directly estimate pitch from spectrogram images. Our new approach demonstrates a very good detection accuracy; a total of 92% of predicted pitch contours have strong or moderate correlations to the true pitch contours. Furthermore, the experimental comparison between our new approach and other state-of-the-art CNN methods reveals that our approach can enhance the detection rate by approximately 5% across various Signal-to-Noise Ratio (SNR) conditions.

*Keywords—pitch detection, F0 extraction, convolutional neural network, machine learning, spectrogram, Smart Mobility*

## I. Introduction

Pitch detection is very widely used for smart mobility features. For example, as shown in Fig.1, pitch contour can be used to train a deep learning neural network for driver's emotion detection, which can alert road rage. Pitch is also a critical feature for speaker identification/verification, particularly useful for children's voice detection for in-vehicle locked children detection. It additionally plays a very important role in vehicle personalized profiles, such as seat position setting through voice or unlocking doors through voice commands. Moreover, vehicle built-in virtual assistants also use pitch for automatic speech recognition because much semantic information is passed on through pitch that is above the phonetic and lexical levels. In tonal languages, for example, Mandarin Chinese, an utterance's relative pitch motion critically contributes to a word's lexical information. Pitch detection has been a popular research topic for many years and is still being investigated today. Traditional pitch detection algorithms include time domain methods and frequency domain methods, for example, the time domain YIN f0 estimator, developed by Alain de Cheveigńe and Hideki Kawahara [2], and the frequency domain F0 estimators [3][4][5]. Previously, statistical approach [14] and HMM (Hidden Markov Model) [15] were also utilized for pitch detection. F0 estimators tailored for specific applications, like detecting musical notes or analyzing speech, are comprehensively grasped, yet their efficacy relies heavily on the data domain. A detector crafted for a specific domain tends to exhibit reduced accuracy when applied to another. Consequently, while numerous F0 estimators are available in the market, only a handful prove suitable across multiple domains. Our method uses spectrogram images directly, so it can avoid challenges of acoustic processing, for example, interpolation artifacts.

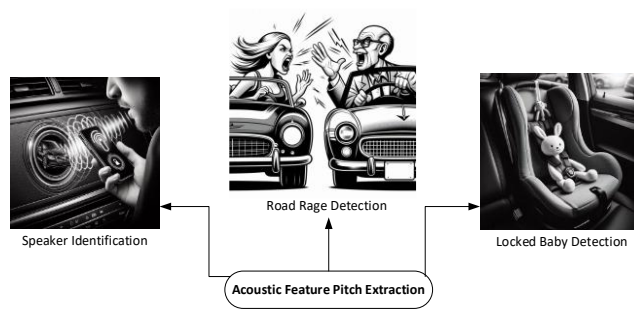

Fig. 1. Pitch Extraction as Technical Enabler for Smart Vehicle Features

In [1], a neural network-based pitch detector was presented based on the human ear's cochlear mechanisms. Lee and Ellis [16] extracted the auto-correlation function features and train a neural network for pitch detection. These previous design initialized our thinking that neural networks can take a set of time/frequency/phase domain data as input and output frequency hypotheses, which can then be translated to pitch. Theoretically, it was a statistical model based on acoustic signal processing. In [6], a CNN was used to classify pitch states, then using GMM (Gaussian Mixture Model) to transfer pitch states to a range of pitch values. We figured out a more straightforward way to extract pitch from spectrogram images through Convolutional Neural Networks (CNN).

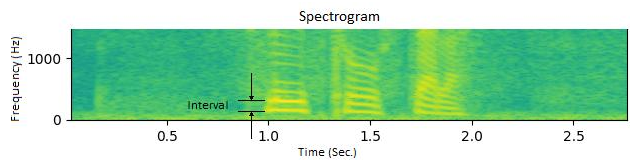

Fig. 2. Pitch Value Indication in Spectrogram Image

As shown in Fig.2, those highlighted bars are called harmonic structure in spectrogram images, and the interval of highlighted bars indicates the F0 value. The bigger the interval, the higher the pitch value. Therefore, the image recognition approach is feasible for extracting F0 values directly. We trained our CNN for regression, which means the output of CNN is real-time pitch values. Therefore, the novelty of our approach includes using harmonic structure in spectrogram to train regression CNN and track pitch continuously.

CNNs are essential tools for machine learning, and CNNs are particularly useful for analyzing image data. For example, CNNs can typically be used to classify images. Besides image classification, a regression layer at the end of the full connection neural network can also be added to predict continuous data. As we know, F0 or fundamental frequency in speech is continuous data, and it plays a very important role in features such as emotion recognition, VAD (Voice Activity Detection), and speech segmentation. As aforementioned, we explore a method to train a CNN model to predict F0 (pitch) through a spectrogram image directly. Compared with the traditional acoustic DSP (Digital Signal Processing) methods, this method can use image enhancement algorithms to highlight pitch bars in a spectrogram to improve robustness. Moreover, to save the computation load, our method can cut the spectrogram as a perfect low-pass filter through the image for pitch detection. The evaluation results are positive and promising, showing good accuracy in real-time testing.

This paper is organized in the following way. First of all, we present the design details of the CNN structure for our new approach in section 2. In Section 3, we give the evaluation results of pitch tracking. Section 4 compares the pitch detection accuracy with the other state-of-the-art approach, the MIT CREPE pitch detector [8]. Section 5 is the general discussion of the advantages and disadvantages of our new approach, and section 6 is the summary and conclusion.

## II. CONVOLUTIONAL NEURAL NETWORK FOR PITCH EXTRACTION

The diagram in Fig.3 illustrates the architecture comprising CNN convolution layers and a fully connected NN (Neural Network) layer. Within this setup, the spectrogram image undergoes filtration to retain solely the highest 2kHz signals. An image buffer that accommodates approximately one second of spectrogram data is utilized, with a processing window duration set at 25 milliseconds. Consequently, the number of NN output nodes totals 44. The input image's pixel dimension is standardized to 27 by 64. To ensure that the convolution scan window captures sufficient information for harmonic bar intervals, it is sized at 16 by 3. Additionally, the pooling layer adopts a window size of 2 by 2. The NN incorporates two hidden layers and is specifically trained to model and predict continuous F0 values.

The deep learning neural network in our system has 500 nodes in the input layer, two hidden layers with 300 nodes and 200 nodes, and, as aforementioned, 44 nodes in the output layer.

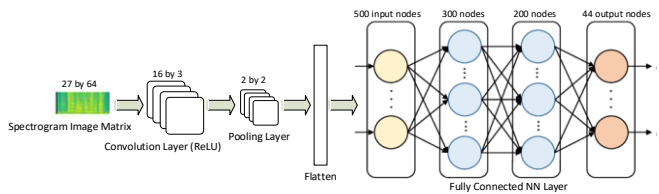

Fig. 3. CNN Structure Diagram for Pitch Detection

Figure 3 shows how spectrogram images were processed to reduce irrelevant background noise and enhance F0 features. Since human pitch values were normally below 1 kHz, a simulated low pass filter on the image was used to remove high-frequency signals. We can cut the image at 1~1.5 kHz; however, we still want to keep enough harmonic structure or highlighted bars for picture convolution, so images from 0 to 2 kHz were retained in this step. Next, the grey scale of these 2 kHz images was tuned to enhance pitch bars, and all other irrelevant background noise was completely removed in this step. The grey scale tuning can enhance the visibility of pitch bars and make pitch-related patterns more pronounced. This step facilitated subsequent analysis and feature extraction. The subplot of GreyScale Tuning in Fig.4 shows we eliminated noise that didn't contribute to pitch analysis, and the resulting images highlighted F0 features without interference from non-pitch-related elements. These enhancements empower researchers and practitioners to extract meaningful insights from spectrogram data, whether in speech analysis, acoustic signal processing, or other applications.

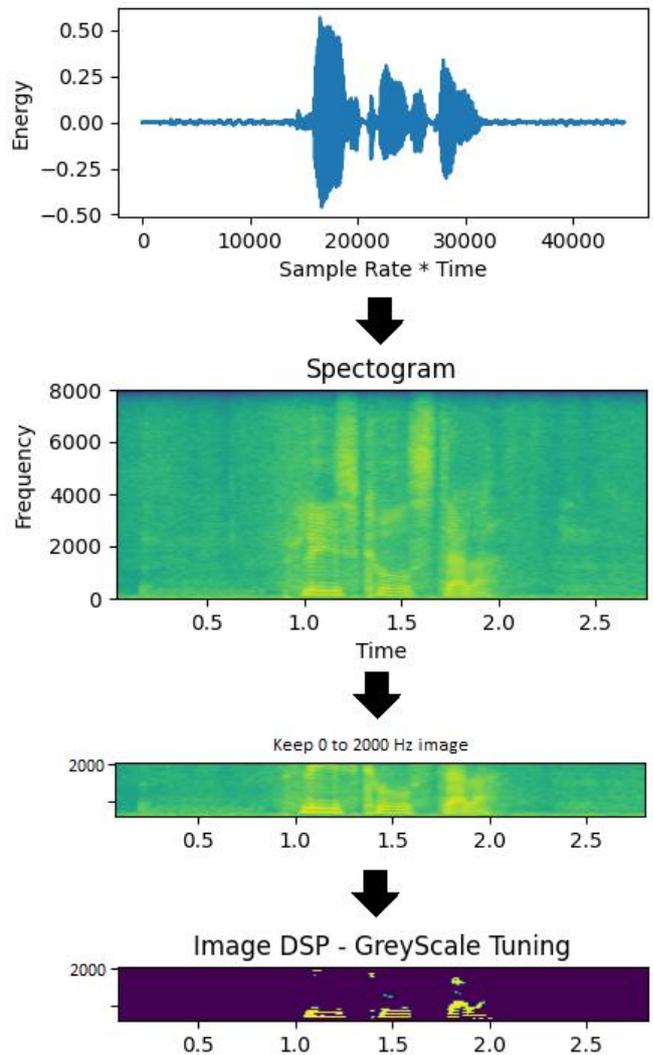

Fig. 4. Spectrogram Image Processing for CNN Training and Testing

## III. EVALUATION RESULTS

We utilized TensorFlow [13] Python Library to implement and evaluate a CNN for pitch detection within the internal research speech data corpus context. This corpus is meticulously curated to be gender balanced and includes voice commands recorded under various driving conditions, featuring a major Signal-to-Noise Ratio (SNR) ranging from 6dB to 20dB. Such diversity in the dataset helps train robust models capable of performing well in noisy environments. Our training dataset comprises approximately 3,000 spectrogram images derived from the audio samples. These spectrograms transform the audio signals into a visual representation that effectively captures the time-dependent frequency spectrum. This makes it suitable for input into CNN models, which excel in extracting patterns from visual data. The CNN's architecture is designed to include multiple convolutional layers that help hierarchically extract features from low-level details, such as edges and gradients, to more abstract features essential for recognizing pitch variations. Each convolutional layer is followed by a pooling layer, which reduces the spatial size of the representation, thus reducing the number of parameters and computations in the network. This is crucial for enhancing the learning efficiency and preventing overfitting, particularly given the limited size of our dataset. The neural network's training process is visualized in Fig.5 through a 'Loss vs. Epoch' graph. This graph is instrumental in monitoring the training progress. It plots the loss metric, calculated as the mean squared error between the network's predictions and the actual data, against the epoch number. Each point on the graph represents the loss after an epoch, clearly indicating how the model's performance evolves over time. Typically, a declining curve indicates learning, whereas a plateau or increase might suggest that the model has begun to overfit or that further training is no longer yielding significant improvements.

For validation, a separate dataset comprising around 800 spectrogram images was used. This dataset was crucial for fine-tuning the CNN model parameters and adjusting the network architecture. The validation process helped ensure that the model performs well on the training data and generalizes effectively to new data, indicating a robust pitch detection system.

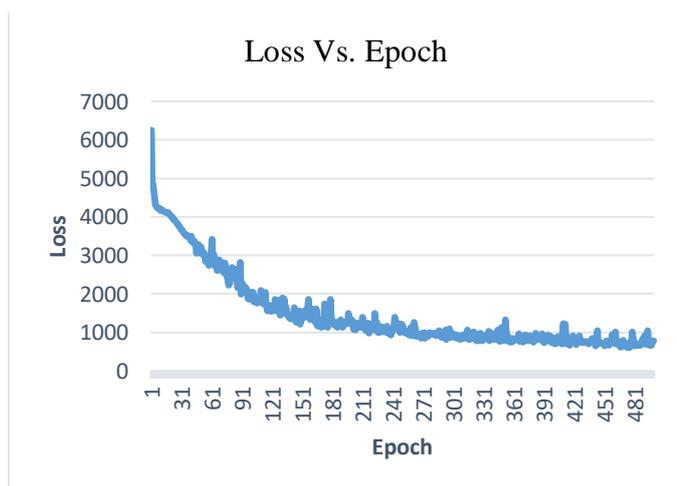

Fig. 5. Loss vs. Epoch Graph

For pitch track testing, another separate dataset comprising around 400 spectrogram images was used, and we used high SNR (around 20dB) speech for this test. Each spectrogram image contains around 30 pitch tracking values, so around 12,000 pitch values were detected and tested. In the evaluation phase of our CNN-based pitch detection system, we employed Pearson's correlation coefficient to quantify the accuracy and reliability of the predictions made by our model. This statistical measure was utilized to assess the degree of linear correlation between two sets of data: the pitch values predicted by our CNN and the actual pitch values, often referred to as the true pitch values. The Pearson correlation coefficient formula is defined as:

$$\rho(X,Y) = cov(X,Y) / \sigma X . \sigma Y \qquad (1)$$

where $cov(X,Y)$ represents the covariance between the predicted and true pitch values, and $\sigma X$ and $\sigma Y$ are the standard deviations of the predicted and true pitch values, respectively.

A higher correlation coefficient indicates a closer approximation of the CNN predictions to the true pitch values. Typically, correlation coefficients greater than 0.7 indicate high correlation, suggesting that the model predictions are very close to the actual data. Coefficients between 0.5 and 0.7 suggest a moderate correlation, still reflecting a reasonable level of predictive accuracy but with some deviation from the true values.

Fig.6 shows an example of correlation evaluation results. The X-axis is the audio frame number, and the Y-axis is the predicted pitch value. Our analysis visualized the results with two figures plotting the true fundamental frequency (F0) values against the CNN predictions. In this example figure, the true F0 values are depicted with a blue line, while the predictions from the CNN are shown with an orange line. The example shows a correlation coefficient of 0.97, indicating an extremely high correlation and demonstrating that the CNN model is proficient in predicting pitch values that closely match the true values.

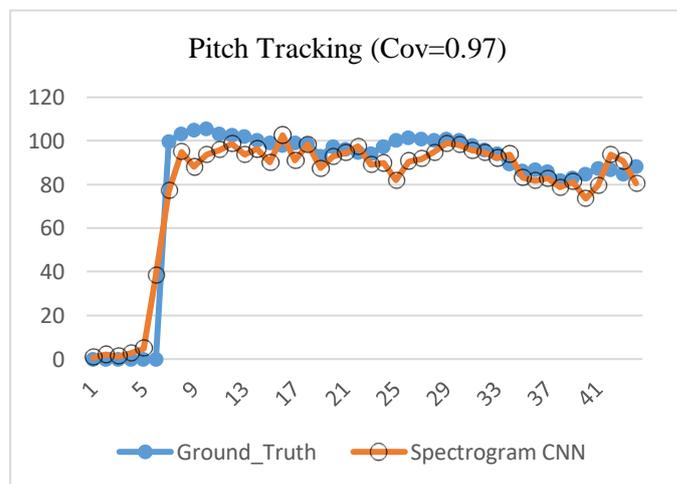

Fig. 6. Example of Using CNN for Continuing F0 Detection, Detected F0 value to each audio frame

From a broader dataset perspective, we analyzed over 12,000 predicted pitch values derived from spectrogram images. Fig.7 shows the distribution of the correlation coefficient. We found that 92% of predicted pitch contours have strong or moderate correlations to the true pitch contours, highlighting the effectiveness of our CNN in accurately capturing pitch nuances in most cases. Of those, 75% show a strong correlation (correlation coefficient ≥ 0.7), and 17% of the predictions fall into the moderate correlation category (correlation coefficient between 0.5 and 0.7), which still supports the utility of our model in practical applications, albeit with some limitations in capturing pitch with absolute precision.

These results underscore our CNN model's capability to process and analyze pitch data with high fidelity and point to areas where further model refinement could enhance prediction accuracy. Using Pearson's correlation coefficient as a metric allows for a clear, quantitative assessment of the model's performance, aligning with established statistical methods to ensure robustness in our evaluation approach.

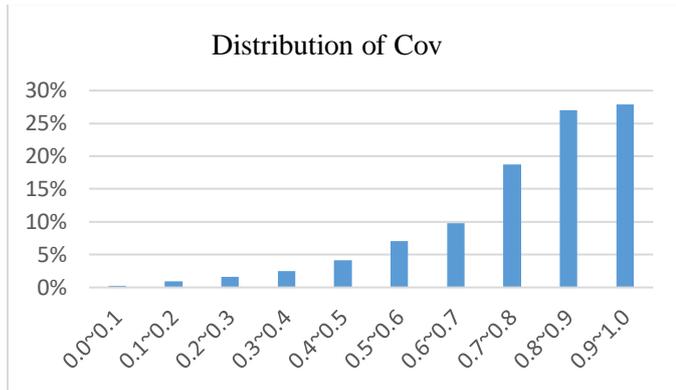

Fig. 7. The Distribution of Correlation Coefficient

## IV. Experiment Comparison with previous AI-based approaches

We compared the performance of our approach with the start-of-art MIT CREPE pitch detector. The CREPE pitch detector uses time domain audio waveforms and CNN to track pitch, and based on their experiments, CREPE outperforms the traditional best performance techniques such as pYIN [2]. Previously, artificial intelligence approaches could not outperform traditional approaches because of a lack of training data[11]. CREPE[8] trained CNN models on a synthetically generated dataset for F0 tracking[12] and achieved state-of-the-art performance results. In this section, we compared our pitch value detection accuracy with CREPE.

In our study, we utilized the same testing dataset as detailed in Section 3. The noise data for this dataset is sourced from road noise recordings in accordance with the ITU-T P.1110 standard [9]. Specifically, our test data includes four distinct types of noise recordings to ensure a comprehensive evaluation of performance under various conditions. These road noise recordings includes (1) Engine idle with the lowest HVAC fan speed, representing minimal ambient noise, (2) City driving at a speed of 60 km/h with the HVAC fan set to medium speed, simulating moderate urban traffic conditions, (3) Highway driving at a speed of 120 km/h with the HVAC fan at its lowest speed, reflecting typical highway noise levels, and (4) Highway driving at the same speed of 120 km/h but with the HVAC fan at medium speed, capturing a higher noise level often experienced on longer journeys.

These varied driving scenarios have been carefully chosen to cover a broad spectrum of Signal-to-Noise Ratios (SNRs), which are crucial for a thorough performance evaluation of our system. By splitting the testing results according to different SNR bins, we can gain deeper insights into our system's performance across a wide range of real-world driving conditions. This detailed analysis allows us to identify strengths and weaknesses in noise management and overall system reliability.

We compare the pitch detection results in terms of detection accuracy rate (AR), and a pitch estimation is considered a correct value if the deviation of the estimated F0 is within 5% of the ground truth F0 value.

$$AR = NC/NP \qquad (2)$$

Where NC is the total number of correct estimations and NP is the number of total pitch frames.

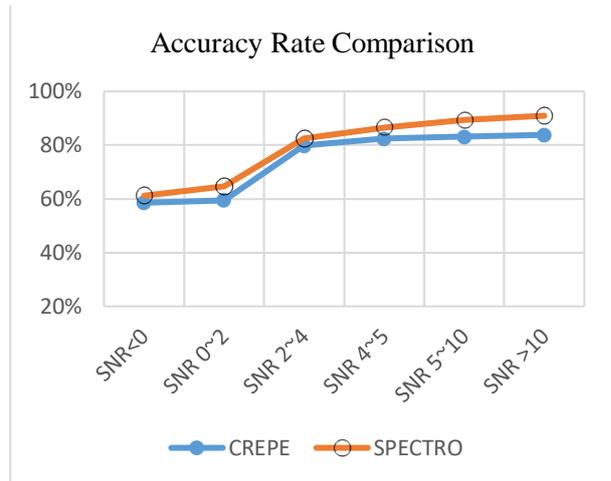

Fig. 8. Detection Rate Comparison with CREPE

Figure 8 illustrates the performance comparison results between two pitch detection approaches. In this figure, "SPECTRO" refers to our pitch detection method, which utilizes spectrograms, while "CREPE" represents the pitch detection system developed by MIT that relies on waveforms.

Our approach demonstrates a significant improvement in accuracy, outperforming the CREPE system by approximately 5% across various Signal-to-Noise Ratio (SNR) conditions. This enhancement is consistent and notable, indicating the robustness of our method in handling different levels of background noise. The comparison clearly highlights the effectiveness of using spectrograms harmonic structure for pitch detection, especially in noisy environments, where maintaining high accuracy is crucial.

Additionally, during our evaluation, we observed that the CREPE pitch detection system had a tendency to over-estimate fundamental frequency (F0) values for certain unvoiced frames. In some instances, CREPE even assigned high F0 values to frames that contained only pure road noise. This could be due to some similar road noise waveforms mis-recognized by CREPE. To remove these over-estimated F0 values, we had to align F0 estimations with confidence scores, which added another layer of post-processing work.

## V. Discussion on Advantages and disadvantages

The estimation of the fundamental frequency has been a long-standing research topic for decades. New technologies and algorithms are being developed and proposed, such as approaches based on deep-learning neural networks [6][7] and pseudo Wigner-Ville distribution [10]. Compared our approaches with two other recently proposed pitch determination algorithms in [6] and [7], our neural network is modelled to solve regression problems. As we mentioned in the introduction, the pitch value is continuous, and the interval of pitch bars indicates actual pitch values. Therefore, we take pitch detection as a regression problem and build a CNN model to calculate continuous pitch value instead of categorizing pitch state. The previous approaches with neural networks took pitch detection as a pitch state classification problem. They quantized continuous pitch range to discrete pitch states, then utilized a neural network to classify pitch states. Those discrete pitch states were tracked to generate a constant pitch contour by maximizing the pitch probability under the temporal continuity constraint of speech [6]. Our approach with convolution neural network outputs pitch values directly without any extract post-processing steps, and it provides another plausible solution for pitch detection.

Secondly, previous AI-based approaches in [6] and [7] utilized an extra step to track pitch contour from pitch state, which increased CPU load and system latency. Our CNN architecture consists of the same number of hidden layers and nodes as the neural network models discussed in references [6] and [7]; as a result, the CPU load and processing time for our CNN model will be comparable to the existing neural network models in [6] and [7]. Still, we eliminate the need to post-process the output values. This streamlined approach reduces complexity, improves efficiency, and minimizes latency.

Everything has two sides, and the spectrogram with the regression CNN approach also has some disadvantages. For example, the regression CNN will create one or two transition values between voice and unvoiced frames. In Fig.6, you can find one transition value on output node 5. These transition values are usually incorrect. Secondly, the system input is changed from one-dimensional time-domain signals to 2D spectrogram images, so the front-end processing computation load will correspondingly increase.

## VI. CONCLUSIONS

The success of CNNs in speech analysis tasks highlights their potential in broader image processing domains. The same principles that enable CNNs to perform well in speech signal processing, such as feature extraction and pattern recognition, can be adapted to other areas where similar challenges exist. This adaptability opens new avenues for applying CNNs to different use cases within the realm of image and signal processing. In the smart mobility area, this CNN-based AI model can extract acoustic features to enable multiple in-vehicle features such as driver's emotion detection, health detection, ID verification, etc. In conclusion, using CNNs for pitch detection provides a robust method for accurately estimating fundamental frequencies and sets a precedent for the broader use of this technology in complex signal processing tasks. As we refine these models and expand their application scopes, the potential for further innovations in speech and general image processing domains remains substantial. This ongoing development of CNN technology promises to deliver even more sophisticated data analysis and interpretation tools, paving the way for advancements across multiple fields of study and industry applications.

In summary, our approach of spectrogram with CNN regression model offers continuous pitch detection while avoiding the post-processing overhead. This efficiency makes it an attractive choice for real-time pitch detection applications.